\documentclass{bulsao}
\usepackage{graphicx}
\usepackage{latexsym}

\begin{document}

\title{GJ\,900: A new hierarchical system with low-mass components}

\author{E.~V. Malogolovets$^1$, Yu.~Yu. Balega$^1$, D.~A. Rastegaev$^1$,
K.-H.~Hofmann$^2$, G.~Weigelt$^2$}

\institute{Special Astrophysical Observatory, RAS, Nizhnii Arkhyz,
Karachai-Cherkessian Republic, 357147 Russia \and Max-Planck
Institut f\"{u}r Radioastronomie, Bonn, Germany}

\offprints{E.~V. Malogolovets, \email{evmag@sao.ru}}

\date{received:January 17, 2007/revised: January 23, 2007}

\titlerunning{GJ\,900: A new hierarchical system with low-mass components}
\authorrunning{Malogolovets et al.}

\abstract{
Speckle interferometric observations made with the 6 m telescope of the Special
Astrophysical Observatory of the Russian Academy of Sciences in 2000 revealed
the triple nature of the nearby ($\pi_{Hip}=51.80\pm1.74$ mas) low-mass young
($\approx200$ Myr) star GJ\,900. The configuration of the triple system allowed it
to be dynamically unstable. Differential photometry performed from 2000 through
2004 yielded $I$- and $K$-band absolute magnitudes and spectral types for the
components to be $I_{A}$=6.66$\pm$0.08, $I_{B}$=9.15$\pm$0.11,
$I_{C}$=10.08$\pm$0.26, $K_{A}$=4.84$\pm$0.08, $K_{B}$=6.76$\pm$0.20,
$K_{C}$=7.39$\pm$0.31, $Sp_{A}$$\approx$K5--K7, $Sp_{B}$$\approx$M3--M4,
$Sp_{C}$$\approx$M5--M6. The ``mass--luminosity'' relation is used to estimate
the individual masses of the components:
$\mathcal{M}_{A}$$\approx$0.64$\mathcal{M}_{\odot}$,
$\mathcal{M}_{B}$$\approx$0.21$\mathcal{M}_{\odot}$,
$\mathcal{M}_{C}$$\approx$0.13$\mathcal{M}_{\odot}$.
%$m_A=0.66$, $m_B=0.21$, $m_C=0.13$ $\mathcal{M}_\odot$ для $I$ полосы и
%$m_A=0.63$, $m_B=0.21$, $m_C=0.14$ $\mathcal{M}_\odot$ для
%$K$ полосы.
From the observations of the components relative motion in the period 2000--2006,
we conclude that GJ\,900 is a hierarchical triple star with the possible
orbital periods P$_{A-BC}$$\approx$80 yrs and P$_{BC}$$\approx$20 yrs. An analysis of the 2MASS
images of the region around GJ\,900 leads us to suggest that the system can
include other very-low-mass components.}

\maketitle

\section{INTRODUCTION}
In the solar neighborhood M-type dwarfs make up about 70\% and no less than
40\% of all population in terms of the number of stars and mass, respectively.
The lifetime of the lowest-luminosity main-sequence stars exceeds the age of
the Universe, making them good candidate objects for the study of the
properties of the Galactic disk including the history of star formation in the
local volume.

The study of young M-type field dwarfs may lead to the discovery of systems
with brown dwarfs. The interest toward the discovery and study of multiple
systems with substellar components has been growing progressively since the
discovery of the first brown dwarf GJ\,229 B (\cite{nakajima95:Malogolovets1_n}). It
is evident that such objects can be found only at small distances from the Sun.
An example of such objects is the group of brown dwarfs GJ\,569 B, which had its
first orbit and dynamical masses estimated from the results of observations
made with three major telescopes: the 6 m Bolshoi Azimuthal Telescope (BTA) of
the Special Astrophysical Observatory of the Russian Academy of Sciences, Keck II, and MMT
(\cite{kenworthy:Malogolovets1_n}).

Until recently, the review by Henry and McCarthy (\cite{henry:Malogolovets1_n})
remained the most representative summary of empirical data on the masses and
luminosities of cool dwarfs obtained using various observational techniques. In
recent years, new data on the principal parameters of low-mass stars in binary
and multiple systems have been published. These data were obtained by combining
different observational methods: adaptive optics imaging combined with accurate
radial-velocity measurements (\cite{segransan00:Malogolovets1_n}),
speckle and
long-baseline interferometry with radial velocities measurements
(\cite{balega07:Malogolovets1_n}; \cite{boden:Malogolovets1_n}), and space
astrometry performed with fine guidance sensors of the Hubble Space Telescope
(\cite{torres:Malogolovets1_n}; \cite{benedict:Malogolovets1_n}). These studies allowed
the main empirical relations to be substantially refined
(\cite{delfosse:Malogolovets1_n}). This concerns, in particular, the
``mass--luminosity'' relation, which is of great importance for the
investigation of parameters of individual stars and of the entire population of
our Galaxy.

In 1998, a speckle interferometric survey of low-mass binaries and suspected
binaries discovered by \emph{Hipparcos} (\cite{esa:Malogolovets1_n}) astrometric satellite was started at
the BTA 6 m telescope (\cite{balega02:Malogolovets1_n}). Observations were
performed at visual and infrared wavelengths with the aim to identify pairs
with fast relative motion of components, which allow model-independent masses
to be determined over a short time interval. In addition to measuring relative
component positions with an accuracy of about 1--2 milliarcseconds (mas), we also
measured the magnitude differences in the $V$, $R$, $I$, $J$, $H$, and
$K$ bands for most of the binaries. Objects of this program include the red
star GJ\,900=Hip\,116384 with the
\emph{Hipparcos} (\cite{esa:Malogolovets1_n}) parallax of
$\pi_{Hip}$=51.80$\pm$1.74 mas. The results of astrometric measurements of this
star allowed it to be suspected as a binary based on a number of indicators
(the object has flag `S' in the \emph{Hipparcos} catalog; see
(\cite{lindegren:Malogolovets1_n})) and that is why we included it into our
program list.

The very first interferometric observations performed in November 2000  with
BTA showed GJ\,900 to be a triple system (\cite{balega06:Malogolovets1_n}). On the reconstructed
image two fainter components are located  $0.5\arcsec$\ and $0.7\arcsec$\ from the main
star. The compact configuration of the GJ\,900 system may indicate that it
belongs to the class of dynamically unstable multiple stars. One must, however,
bear in mind that the components of this system may appear to be located
at a comparable distances from the main star only in the sky-plane projection.
To verify these assumptions, we included this system into the program of
monitoring of the relative motion of components.

In 2002--2003, Martin (\cite{martin:Malogolovets1_n}) observed GJ\,900 with
the CIAO adaptive optics system of the 8.2 m Subaru Telescope and confirmed the two faint companions
of the central object. The results of two observations in the infrared bands
$H$ and $K$ separated only by a five-month interval led him to conclude that the
system is dynamically bound.

In this paper we report the results of the interferometric
measurements of positional parameters and magnitude differences
of the components of GJ\,900 made during the period from
November, 2000 through December, 2006, determine the absolute
magnitudes and estimate the masses of the stars. We analyze the
possible dynamical stability of the system based on the
observations performed.

\section{OBSERVATIONS AND DATA ANALYSIS}
Speckle interferometric observations of GJ\,900 were made with the BTA 6 m
telescope in the $V$ ($\lambda/\Delta\lambda$=550/30 nm), $I$ ($\lambda/\Delta\lambda$=800/100 nm),
and $K$($\lambda/\Delta\lambda$=2115/214 nm) bands, where $\lambda$
is the central wavelenght, $\Delta\lambda$ is the half-width of the band.
A fast 512$\times$512 Sony ICX085
CCD combined with a three-stage image intensifier was used as a detector from 2000
through 2004. During our 2006 observations we employed a new EMCCD
(Electron Multiplying Charge Coupled Device) system with
higher quantum efficiency and linearity. The image scale was
equal to 4.1 and 6.7 mas/pixel for the first and second facilities,
respectively. The exposure time of speckle interferograms varied from 5 to
20 msec depending on brightness of an object and seeing conditions. Infrared
observations were made with HAWAII infrared detector of the
Max-Planck-Institute for Radio Astronomy.
 Table\ref{tab1:Malogolovets1_n} gives a log of observations.
 This table gives the following data for each measurement: date as a
fraction of Besselian year, seeing $\beta$ in arcseconds, the number of
speckle interferograms in each series, and the name of the filter.

\begin{table}[tbp]
\begin{center}
\caption{Log of observations} \label{tab1:Malogolovets1_n}
\bigskip
\begin{tabular}{c|c|c|c}
\hline
Date      &  $\beta$    & N       & Filter \\[-5pt]
	  &  arcsec     &         &                       \\
\hline
2000.8754  &  1.5        & 900     & I                 \\
2003.7880  &  1          & 1000    & K                \\
2003.9248  &  1.5        & 2000    & I                 \\
2004.8208  &  1          & 2000    & I                 \\
2006.9465  &  1          & 2000    & V                 \\
2006.9465  &  1          & 2000    & I                 \\
\hline
\end{tabular}
\end{center}
\end{table}

\begin{figure}[t]
\begin{center}
\includegraphics[width=6cm]{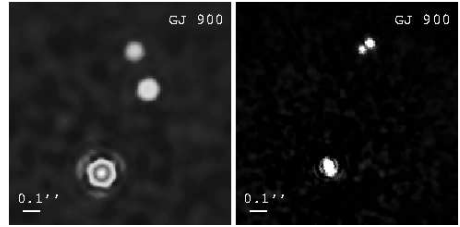}%[scale=0.8]
\caption{
Images of GJ\,900 reconstructed from the 2115/214-nm
filter observations with the BTA 6 m telescope in August 2003 (left) and
from the 800/100-nm in December 2006 (right). North is at the top and East
on the left.}
\label{fig1:Malogolovets1_n}
\end{center}
\end{figure}

The angular distances $\rho$, position angles $\theta$, and magnitude
differences between the components $\Delta$m from speckle interferometric
observation with the BTA telescope are given in Table 2. Due to low
signal-to-noise ratio in the 2006.9465 measurements in $V$-band, the
relative positions for this data are not presented in the table.
A description of the technique of the determination of relative positions and
component magnitude differences inferred from the averaged over the series
power spectra of speckle interferograms can be found in the paper by
Balega et al. (\cite{balega02:Malogolovets1_n}). The resolution diffraction limit
was equal to $0.022\arcsec$, $0.033\arcsec$, and $0.088\arcsec$ in the $V$, $I$,
and $K$ bands, respectively. The accuracy of the measurement of position
parameters is equal to 0.3--1.0$^{\circ}$ and 3--8~mas in position angle
$\theta$ and angular separation $\rho$, respectively. The errors of measured
$\theta$ and $\rho$ depend on a number of parameters: component separation,
magnitude differences, and seeing $\beta$. The accuracy of the determination of
component magnitude differences from the reconstructed power spectra is also a
function of the same parameters. This accuracy varies from 0.05 to 0.2 for
objects with $m_{V}$=8--10. The modulus of the Fourier transform of the object
(visibility) was obtained from the series of speckle interferograms with the classical
speckle interferometry method. For image reconstruction we used the
bispectrum speckle interferometry method
(\cite{weigelt:Malogolovets1_n}; \cite{lohm:Malogolovets1_n}).
Fig.~\ref{fig1:Malogolovets1_n} shows the
reconstructed infrared and visual images of GJ\,900 based on observations made
in 2003 and 2006.

\begin{figure*}[t]
\begin{center}
\includegraphics[width=12cm]{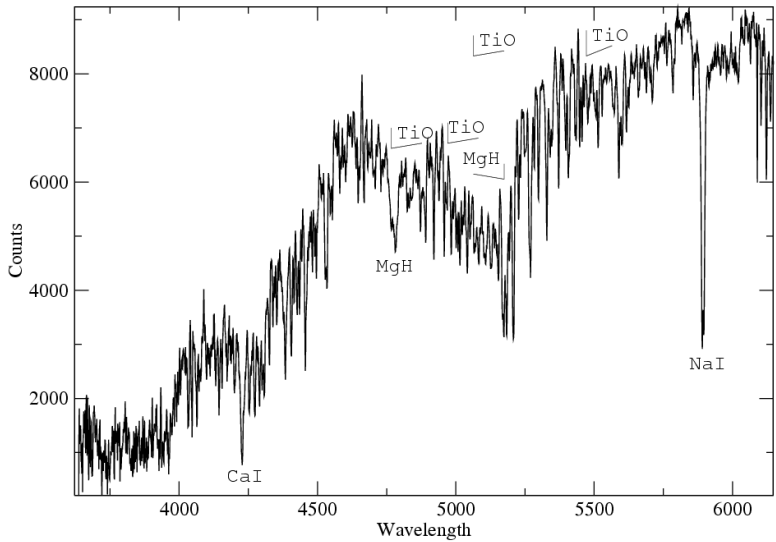}%[scale=0.8]
\caption{Spectrum of GJ\,900 in the 3600 --
6200 \AA\ wavelength interval taken with the UAGS spectrograph of
Zeiss-1000 telescope in October 2006. The strongest absorption
lines and bands --- those of TiO, MgH, CaI, and NaI --- are
indicated.}
\label{fig2:Malogolovets1_n}
\end{center}
\end{figure*}

To specify the spectral type of the object, the
3600 -- 6200 \AA\ spectrum of the object
was taken in October 2006 with the UAGS spectrograph
of the Zeiss-1000 telescope with a dispersion of 1.35 \AA/pixel \ref{fig2:Malogolovets1_n}.

\section{ABSOLUTE MAGNITUDES, MASSES AND SPECTRAL TYPES OF THE COMPONENTS}

To compute the $I$- and $K$-band absolute magnitudes of the components of
GJ\,900, we used the results of differential speckle photometry performed with
the BTA telescope (Table 2) and published integrated magnitudes of the system
in these bands. For the sake of completeness,  this table also includes the
photometric data from (\cite{martin:Malogolovets1_n}). The component magnitude differences
determined using the results of speckle interferometry performed with BTA
agree, on the whole, with those inferred from observations performed
with the Subaru Telescope adaptive optics. At the same time, the results of
differential photometry do not rule out the variability of one or several
components of the system. The integrated $I$-band magnitude determined from the
color index of the system, $V-I=1.65$ (\cite{bessel:Malogolovets1_n}), and the
visual magnitude adopted from (\cite{rNLTT:Malogolovets1_n}) is equal to
$m_{I}$=7.94. The integrated  $K$-band magnitude is equal to
$m_{K}$=6.01$\pm$0.01 (\cite{alonso:Malogolovets1_n}). We combine these data
with the \emph{Hipparcos} parallax and magnitude differences from Table 2 to
infer the following estimates for the absolute magnitudes of the components of
GJ\,900:

$I_{A}$=6.66$\pm$0.08, $K_{A}$=4.84$\pm$0.08,

$I_{B}$=9.15$\pm$0.11, $K_{B}$=6.76$\pm$0.20,

$I_{C}$=10.08$\pm$0.26, $K_{C}$=7.39$\pm$0.31,

where
for the $I$ band we used the mean result averaged over two observations made
with the BTA 6 m telescope: $\Delta m_{I}^{AB}$=2.49$\pm$0.07, and $\Delta
m_{I}^{AC}=3.42\pm0.24$. The $H$-band component magnitude differences measured
using adaptive optics (\cite{martin:Malogolovets1_n}) yield the following
absolute magnitudes for the components:

$H_{A}$=5.11$\pm$0.08,

$H_{B}$=6.85$\pm$0.09,

$H_{C}$=7.54$\pm$0.15.

\begin{table*}[tbp]
\begin{center}
\caption{Speckle interferometric and adaptive optics measurements of GJ\,900}
\label{dm:Malogolovets1_n}
\bigskip
\begin{tabular}{c|c|c|c|c|c|c|c|c|c}
\hline
 Date,     & Component    & $\rho$  & $\sigma_{\rho}$ & $\theta^{\circ}$ & $\sigma_{\theta}$ & $\Delta m$ & $\sigma_{\Delta m}$ & Filter & Reference \\ [-5pt]
 BY        & vector       & mas     & mas             &                  &                   &           &                     &         &        \\
\hline
 2000.8754 & AB       &  417     & 3             & 316.4          & 0.4              & 2.42      & 0.15                & $I$                   & (\cite{balega06:Malogolovets1_n})  \\
       & AC       &  716     & 5             & 344.3          & 0.4              & 3.65      & 0.22                &                           &                  \\
       & BC       &  399     & 6             & 13.6           & 0.6              &           &                     &                           &                  \\

 2002.5990 & AB       &  510     & 10            & 324.5          & 0.1              & 1.78      & 0.02                & $H$                   & (\cite{martin:Malogolovets1_n}) \\
       & AC       &  760     & 10            & 344.0          & 0.1              & 2.55      & 0.03                &                       &               \\

 2002.5990 & AB       &  510     & 10            & 324.5          & 0.1              & 1.61      & 0.03                & $K$                   & (\cite{martin:Malogolovets1_n})  \\
       & AC       &  760     & 10            & 344.0          & 0.1              & 2.38      & 0.04                &                       &               \\

 2003.0480 & AB       &  520     & 20            & 327.4          & 0.1              & 1.70      & 0.04                & $H$                   & (\cite{martin:Malogolovets1_n}) \\
       & AC       &  740     & 20            & 343.9          & 0.1              & 2.31      & 0.06                &                       &               \\

 2003.7880 & AB       &  557     & 5             & 331.3          & 0.6              & 1.92      & 0.18                & $K$                  & This paper    \\
%&&&&& \\
       & AC       &  733     & 6             & 345.1          & 0.6              & 2.55      & 0.30                &                           &               \\
       & BC       &  234     & 4             & 19.9           & 1.0              & 0.63      & 0.35                &                           &               \\

 2003.9248 & AB       &  559     & 4             & 331.7          & 0.4              &           &                     & $I$                   & This paper    \\
%&&&&&paper \\
       & AC       &  726     & 5             & 345.1          & 0.4              &           &                     &                           &               \\
       & BC       &  224     & 6             & 20.4           & 0.6              &           &                     &                           &               \\

 2004.8208 & AB       &  606     & 3             & 335.8          & 0.3              & 2.56      & 0.06                & $I$                   & This paper    \\
       & AC       &  714     & 4             & 345.5          & 0.4              & 3.18      & 0.22                &                           &               \\
       & BC       &  155     & 5             & 26.7           & 0.5              &           &                     &                           &               \\

 2006.9465 & AB       &  751     & 3             & 342.5          & 0.3              &           &                     & $I$                   & This paper    \\
%&&&&&paper \\
       & AC       &  708     & 8             & 344.7          & 0.7              &           &                     &                           &               \\
       & BC       &  51      & 9             & 130.3          & 0.8              &           &                     &                           &               \\
\hline
\end{tabular}
\end{center}
\end{table*}

The luminosities of the components allow their masses to be
estimated from the ``mass--luminosity'' relation, however, to do
this, we must know the age and metallicity of the system.
%На молодость системы указывает повышенная эмиссия в линиях H$_{\alpha}$,
%Ca II H и K (Стайфер и Хартманн, 1986; Бопп, 1987; Гиампапа и др., 1989).

Gizis et al. (\cite{gizis:Malogolovets1_n}) compared the activity of a large
sample of nearby M-type field dwarfs with the activity of M-type dwarfs in open
clusters and calibrated the
``age--activity'' relation. We use the
 $V-I = -6.91+1.05log(\tau)$ relation to infer an age of 10$^8$~Myr for the adopted
distance modulus of $m$--$M$=1.43. X-ray luminosity of late-type stars is yet
another indicator of their activity. According to $ROSAT$ observations, the
X-ray luminosity of GJ\,900 is equal to $L_{x}=108.8\times10^{27}$~erg/s
(\cite{huensch:Malogolovets1_n}). This value corresponds to the M-dwarf age of
100--200 Myr.

The fluxes in the CaII H and K and MgII h and k lines are good indicators of
stellar activity and hence of stellar age
(\cite{soderblom:Malogolovets1_n}; \cite{wilson:Malogolovets1_n}). However, no relations
for age as a function of these indicators have been derived for stars of
spectral types later than K. The only solution is to compare  GJ\,900 with
stars having similar spectra with the age determined using other methods.
For GJ\,900 Giampapa et al. (\cite{giampapa:Malogolovets1_n}) report CaII H and K line flux
measurements: $log(F_{CaII})=5.18$ erg/s$\times$cm$^2$. We adopted the
MgII-line flux $log(F_{MgII})=5.83$ erg/s$\times$cm$^2$ from
(\cite{panagi:Malogolovets1_n}).
For comparison, we selected two stars of similar
spectral types: GJ\,212 and GJ\,879. The CaII H and K line fluxes of these stars are
equal to $log(F_{CaII})=5.49$ erg/s$\times$cm$^2$
(\cite{stauffer:Malogolovets1_n}; \cite{rutten89:Malogolovets1_n}) and
$log(F_{CaII})=5.97$erg/s$\times$cm$^2$
(\cite{robinson:Malogolovets1_n}; \cite{rutten91:Malogolovets1_n}),
respectively. The MgII h and k line fluxes are equal to
$log(F_{MgII})=5.36$ erg/s$\times$cm$^2$
(\cite{panagi:Malogolovets1_n}) and $log(F_{MgII})=5.98$erg/s$\times$cm$^2$,
respectively (\cite{panagi:Malogolovets1_n}). The ages of these stars were
inferred from kinematics, isochrones, and lithium abundance and vary from
100 to 200 Myr.

Zuckerman et al. (\cite{zuckerman:Malogolovets1_n}) believe GJ\,900 to be a
possible member of the Carina-Near moving group with an estimated age of 200
$\pm$ 50 Myr. However, the star's membership in this group is doubtable. The
radial velocity of GJ\,900 is equal to \mbox{--10 km/s} and differs strongly
from that of the core of the group (+20 km/s (\cite{zuckerman:Malogolovets1_n})).
The center of the moving group is at a distance of 30--50 pc from us, whereas
GJ\,900 is located within mere 19 pc from the Sun. The equivalent width of the
LiI $\lambda$6708 line (\cite{zboril:Malogolovets1_n}) in the spectrum of GJ\,900
is several factors of ten smaller than in that of the corresponding lines in
the spectra of the main members of the group.

Martin (\cite{martin:Malogolovets1_n}) used the results of the kinematical survey
of Montes et al. (\cite{montes:Malogolovets1_n}) to estimate the age of the star
at 50--100 Myr.

Thus all the available observational data indicate that GJ\,900 is a young
system with the age of about 200$\pm$100 Myr.

Since GJ\,900 is located in the immediate proximity to the Sun (d=19.3 pc)
and belongs to the galactic disk population, we can suppose that its metal
abundance is close to the Sun's value. This assumption is supported by the
results of spectroscopic study of the system performed by Zboril and
Byrne (\cite{zboril:Malogolovets1_n}). They give the metallicity of the star
[M/H]=-0.1$\pm$0.2.

Zboril and Byrne (\cite{zboril:Malogolovets1_n}) determined the effective
temperature of GJ\,900 from sensitive photospheric lines and molecular bands
with the allowance for surface gravity and microturbulence. Their estimate,
T$_{eff}$=4000\,K, is lower by 200\,K than the temperature inferred from the
$B$--$V$ and \mbox{$R$--$I$} color indices (\cite{gliese:Malogolovets1_n}). It is
evident that this temperature and the corresponding spectral type K7 refer to the
main component of the system.

\begin{figure*}[]
\begin{center}
\includegraphics[width=16cm]{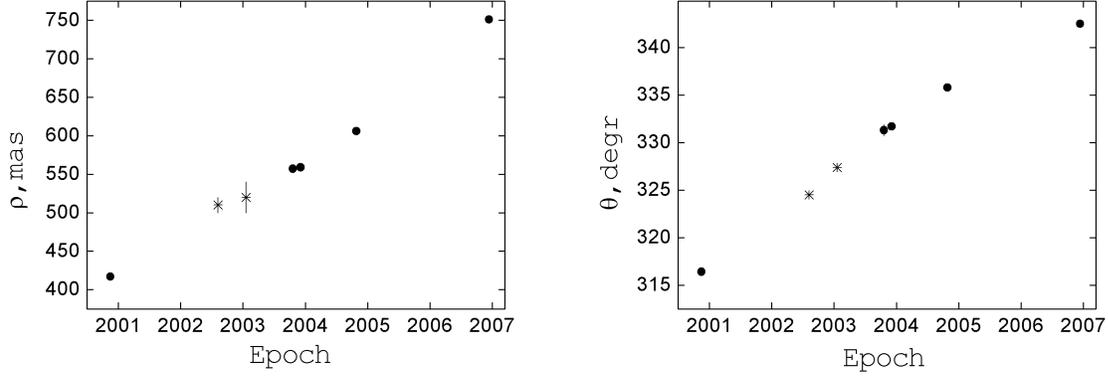}%[scale=0.1]
\caption{Measurement of the angular separation and
position angle of GJ\,900 B relative of GJ\,900 A.
Circles and asterisks show the
speckle interferometric measurements made with the BTA 6 m telescope
and the measurements made with the Subaru Telescope (\cite{martin:Malogolovets1_n}) using
adaptive optics, respectively. The bars show the measurement errors.}
\label{fig3:Malogolovets1_n}
\end{center}
\end{figure*}

\begin{figure*}[]
\begin{center}
\includegraphics[width=16cm]{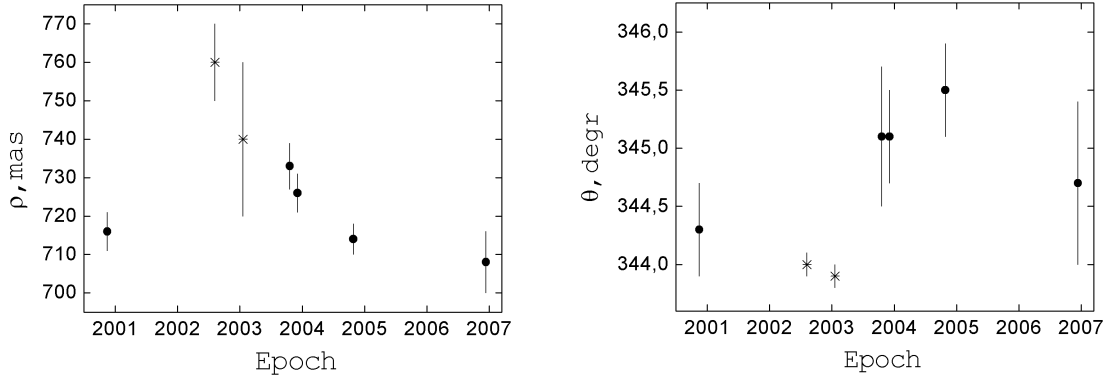} %[%scale=0.1]
\caption{Same as Fig.~3, but for GJ\,900 C.}
\label{fig4:Malogolovets1_n}
\end{center}
\end{figure*}

\begin{figure*}[]
\begin{center}
\includegraphics[width=5cm]{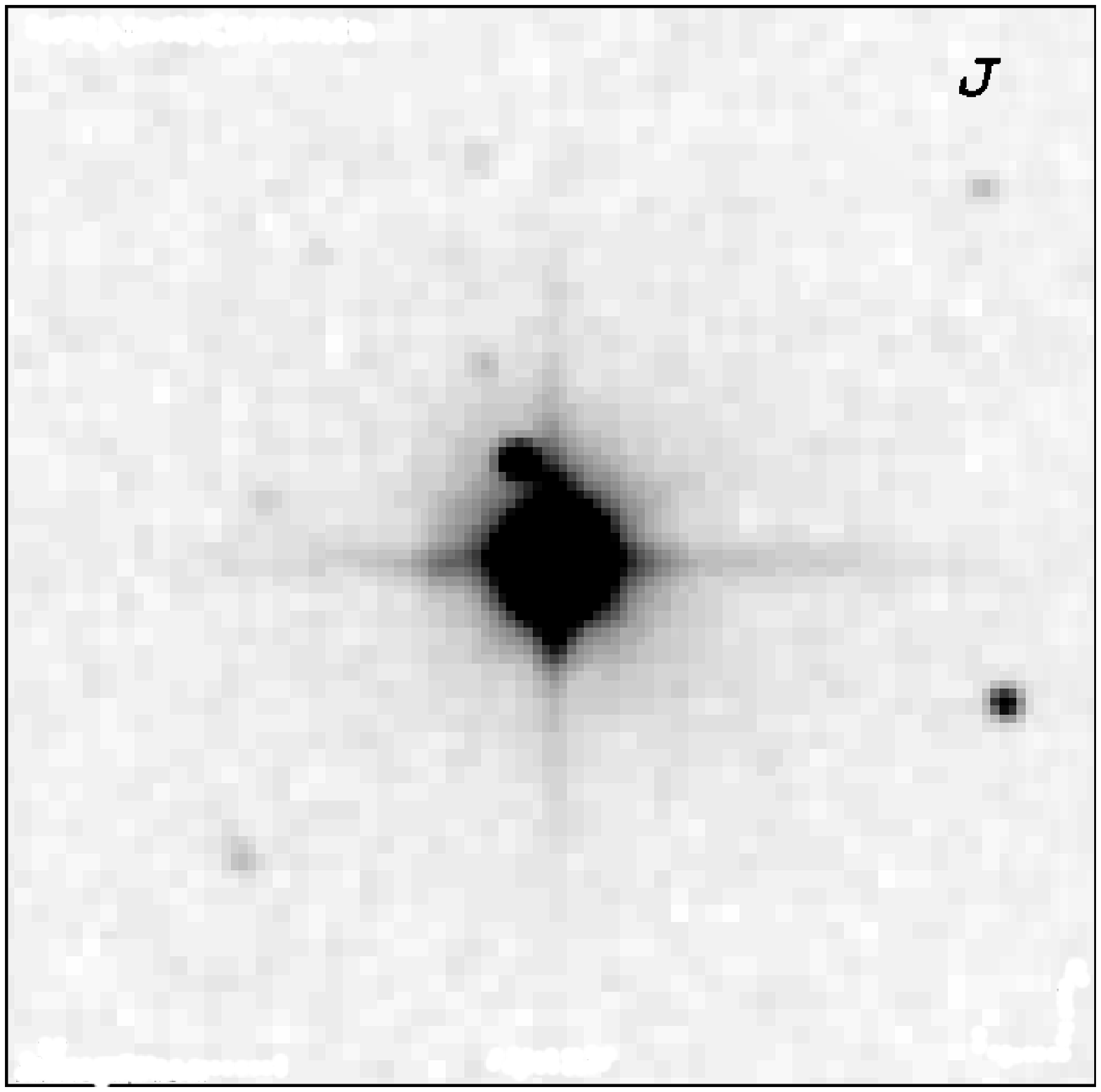}%[scale=0.3]
\includegraphics[width=5cm]{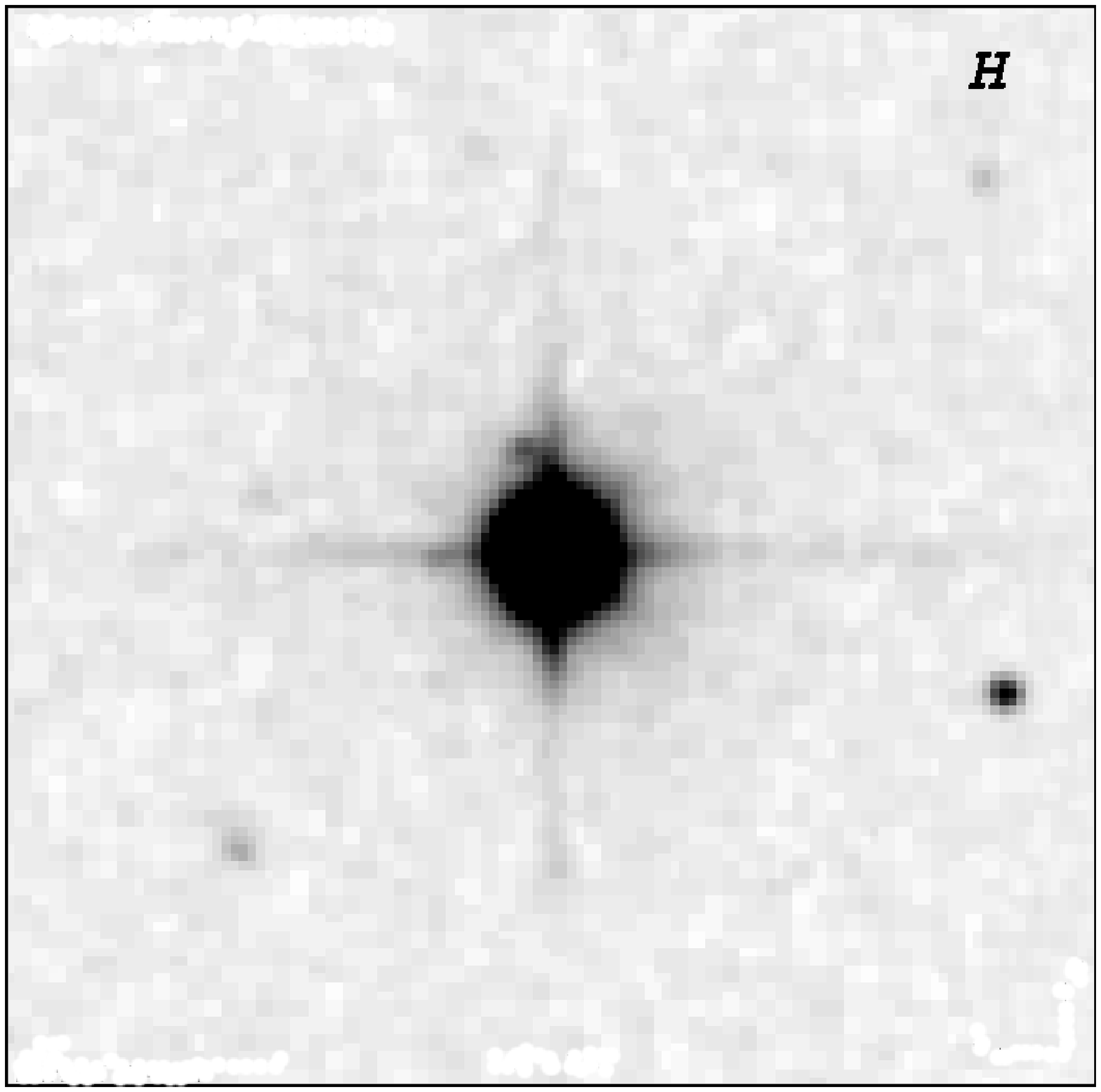}%[scale=0.3]
\includegraphics[width=5cm]{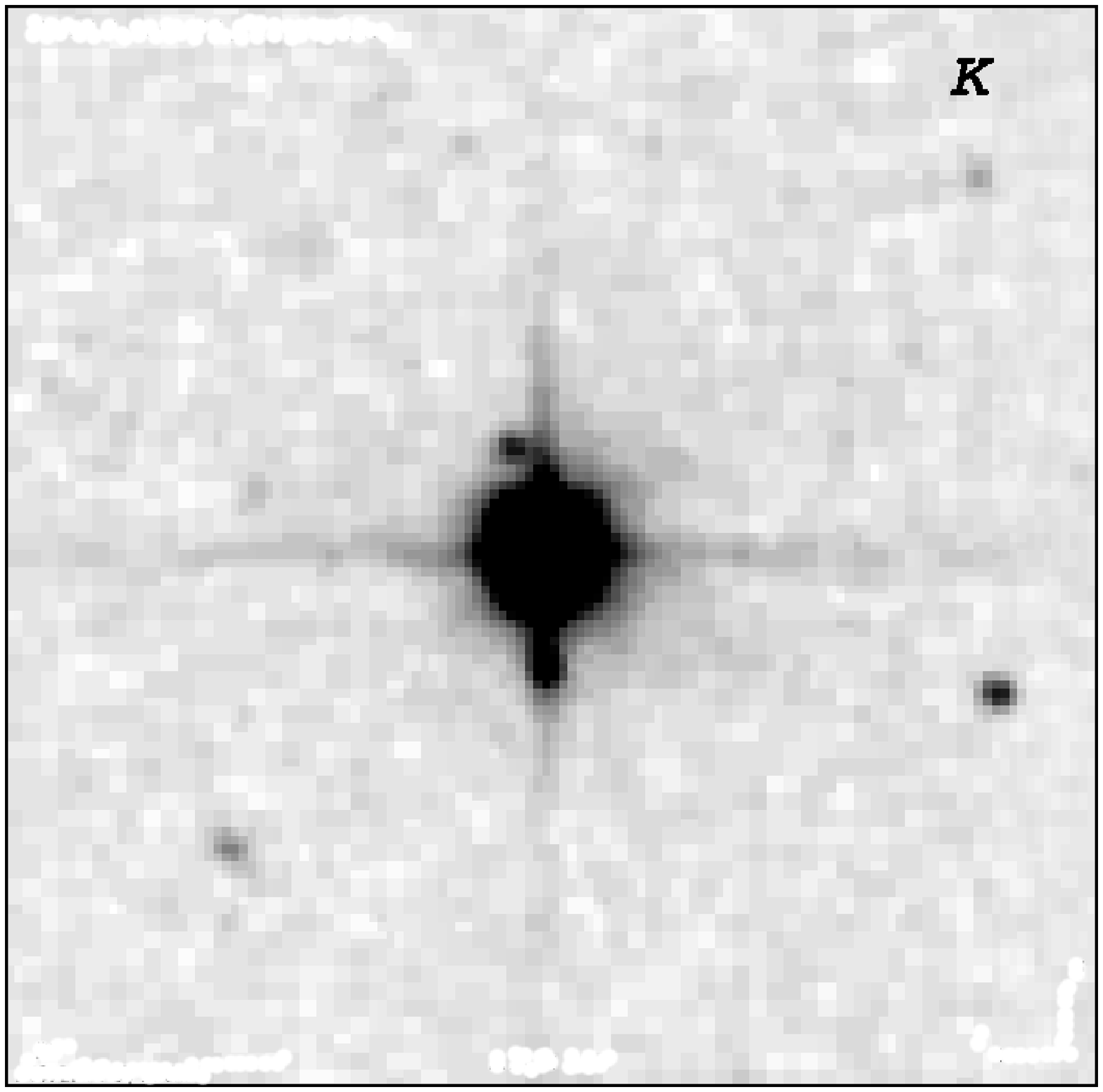}%[scale=0.3]
\caption{$J$, $H$, and $K$-band 2MASS images of the
2.1$'\times$2.1$'$ region around GJ\,900 (\cite{skrutskie:Malogolovets1_n}).
North is at the top and East on the left.} \label{fig5:Malogolovets1_n}
\end{center}
\end{figure*}
Speckle interferometric measurements of the $I-,$ $K$-band component
magnitude differences and adaptive optics measurements in the  $H$ and
$K$ bands allow the effective temperature and hence the spectral type of the
primary component to be estimated using the calibrated relation between
temperature and  $V$--$I$ and $V$--$K$ color indices
(\cite{alonso96:Malogolovets1_n}). To estimate the $V$-band luminosity of the
primary component, we used the $I$, $H$, and $K$-band absolute magnitudes and
theoretical isochrones for the age of 200 Myr (\cite{baraffe:Malogolovets1_n}).
The mean $V_{A}$ inferred from three isochrones is equal to 8.35$\pm$0.07. In
this case, given the distance modulus of  $m-M=1.43$, the color indices of the
primary should be equal to ($V$--$I$)$_{A}$=1.70$\pm$0.08 and
($V$--$K$)$_{A}$=3.52$\pm$0.09. The calibration of temperature in terms of color
index derived by Alonso et al. (\cite{alonso96:Malogolovets1_n}) depends only
slightly on the metallicity of the star. We assume that iron abundance is equal
to the solar value to infer from the estimated ($V$--$I$)$_{A}$ and ($V$--$K$)$_{A}$
indices the temperature of component GJ\,900 A,
 T$_{eff}^{A}$=4079$\pm$180 K, which corresponds to a late K-type dwarf.

An analysis of the GJ\,900 spectrum obtained with the UAGS spectrograph on the
Zeiss-1000 telescope showed that
the energy distribution and relative intensities of single strong lines
correspond to those of a  K5--K7-type star (Fig.
\ref{fig2:Malogolovets1_n}). The characteristic features of the spectrum
include strong TiO and MgH absorption bands.

For the age of 100--200 Myr and solar chemical composition the
$M_I-\mathcal{M}$ and $M_K-\mathcal{M}$ evolutionary tracks for low-mass stars
computed by Baraffe et al. (\cite{baraffe:Malogolovets1_n}) imply a
primary-component mass in the interval from 0.64 to 0.67 $\mathcal{M}_{\odot}$.
This mass agrees best with the primary spectral type of K5 -- K7. The masses of
lower-mass components B and C inferred from the same tracks show a much greater
scatter. The mass of GJ\,900 B is estimated to range from 0.28 to 0.34
$\mathcal{M}_{\odot}$, and that of GJ\,900 C, from 0.16 to 0.24
$\mathcal{M}_{\odot}$. Note the scatter of possible masses is even greater when
estimated from the  $K$-band photometry of the stars studied. The resulting
mass estimates correspond to the spectral types of  M3 -- M4 and M5 -- M6 for
the second and third components, respectively.

\section{RELATIVE MOTION OF COMPONENTS, LIKELY ORBITAL PERIODS, AND THE DYNAMICAL STABILITY
OF THE SYSTEM}

The proper motion of GJ\,900 is equal to 344 mas/yr. If we were dealing with
accidental projection, components B and C would have shifted by almost 2$\arcsec$
relative to component A during the observing period from 2000 to 2006. However,
the mutual positions of these components changed insignificantly during our
monitoring program (Fig. \ref{fig3:Malogolovets1_n} and \ref{fig4:Malogolovets1_n}).
The average annual variation in the positional
parameters of component B relative to component A was equal to  4.3$^{\circ}$
and 55 mas in position angle and angular separation, respectively. The
variation of the position of component C relative to component B is equal to
19.2$^{\circ}$/yr and 57 mas/yr for $\theta$ and $\rho$, respectively.
It follows from the relative position changes that the components B and C
form the inner short-period subsystem, which moves with the component A around
the GJ\,900 mass center. The average annual variations of the
positional parameters imply an orbital period of about 80 and 20 years for the
subsystems  A-BC and BC, respectively. Thus GJ\,900 is a gravitationally bound
hierarchical multiple system. The planes of the orbital motion of components
in the subsystem BC and of component
A are tilted most probably at a large angle to each other, resulting in the observed
configuration.

\section{THE PRESENCE OF EXTRA COMPONENTS IN THE SYSTEM}

To find the eventual faint components in the GJ\,900 system, we analyzed the
2MASS images (\cite{skrutskie:Malogolovets1_n}) taken in August, 2000. We found
on the  $J$-, $H$-, and $K$-band images a faint companion of $\approx$12--13
magnitude $\approx12\arcsec$\ Northeast of the central object
(Fig.\ref{fig5:Malogolovets1_n}). The $K$-band image also shows another component,
at $\approx15\arcsec$\ South of the central source. The probability
for a star to be located accidentally within the 30-arcsec field in the region
studied is equal to about one percent. Hence GJ\,900 is very likely to be a
quadruple or even a  quintuple system. From the intensity ratio it follows
that the faint components are late M dwarfs.

To verify these hypotheses, $I$-band images of the GJ\,900 vicinity were taken
by A.V. Moiseev in February 2007 with the SCORPIO focal reducer on the BTA
telescope. We took and then averaged 25 ten-second exposures. The
limiting magnitude of the resulting image is 17$^m$, but we found no objects at
the location of the Northeastern component. This result suggests that either
this component is too red to be seen in the $I$ band --- and this assumption is
supported by the absence of the object on the $I$-band plate of the DSS2 survey
--- or that, because of the proper motion of the component or that of GJ\,900, this
object is now projected onto GJ\,900. Hence infrared photometry is needed to
make the final conclusion about the membership of components in the multiple
system. If the components found do not belong to the multiple system GJ\,900,
they should shift considerably relative to GJ\,900 since the time when the
corresponding 2MASS images were taken. If the relative positions of these
components remain unchanged, this object would be a unique low-mass multiple
system, which is of interest for testing the theory of star formation and
dynamical evolution of stars.

\section{CONCLUSIONS}
Speckle interferometric observations made with BTA 6 m telescope
during the period
from 2000 to 2006 showed that  GJ\,900 is a gravitationally bound triple star.
This system belongs to the population of the thin disk of the Galaxy and has an
age of 200$\pm$100 Myr. The absolute magnitudes of the components are:
$I_{A}$=6.66$\pm$0.08, $I_{B}$=9.15$\pm$0.11, $I_{C}$=10.08$\pm$0.26,
$K_{A}$=4.84$\pm$0.08, $K_{B}$=6.76$\pm$0.20, and $K_{C}$=7.39$\pm$0.31 and
correspond to the spectral types of $Sp_{A}$$\approx$K5--K7, $Sp_{B}$$\approx$M3--M4,
and $Sp_{C}$$\approx$M5--M6. We used the evolutionary tracks of Baraffe et al.
(\cite{baraffe:Malogolovets1_n}) to compute the component masses for the solar
metallicity and for the age of  200$\pm$100 Myr:
$\mathcal{M}_{A}$$\approx$0.64--0.67$\mathcal{M}_{\odot}$,
$\mathcal{M}_{B}$$\approx$0.28--0.34$\mathcal{M}_{\odot}$,
$\mathcal{M}_{C}$$\approx$0.16--0.24$\mathcal{M}_{\odot}$. The estimated masses
and absolute magnitudes of the components agree with the results obtained in
the $H$ and $K$ bands with the Subaru Telescope adaptive optics
(\cite{martin:Malogolovets1_n}).

We estimated the  orbital periods of the components, $P_{BC}\approx$20 yr and
$P_{A-BC}\approx$80 yr, and concluded that  GJ\,900 is an hierarchical
multiple star. We explain the comparable mutual angular separations between the
components by the effect of sky-plane projection.

We examined the $J$-, $H$-, and $K$-band images of the 2MASS survey and found
possible distant components of the GJ\,900 system at 12$\arcsec$ and 15$\arcsec$.
If these components are
gravitationally bound to the triple star, GJ\,900 should be a young quintuple
system of red dwarfs.

\begin{acknowledgements}
We are grateful to A.~Burenkov
for taking and reducing a spectrum of  GJ\,900 with the Zeiss-1000 telescope
as well as to A.~Moiseev for taking direct images of
the object studied with the BTA 6 m telescope. This research has made use
of the SIMBAD database, operated at CDS, Strasbourg, France. We acknowledge
support from Russian Foundation for Basic Research through research grant
07-02-01489.
\end{acknowledgements}

\end{document}